# Generalized bioinspired approach to a daytime radiative cooling "skin"

*Meng Yang, Weizhi Zou, Jing Guo, Zhenchao Qian, Heng Luo, Shijia Yang, Ning Zhao\*, Lorenzo Pattelli\*, Jian Xu\*, Diederik S. Wiersma*

M. Yang, W. Zou, J. Guo, Z. Qian, H. Luo, S. Yang, N. Zhao
Beijing National Laboratory for Molecular Sciences, CAS Research/Education Center for Excellence in Molecular Sciences, Laboratory of Polymer Physics and Chemistry, Institute of Chemistry, Chinese Academy of Sciences, Beijing 100190, P. R. China
E-mail: zhaoning@iccas.ac.cn

M. Yang, W. Zou, J. Guo, Z. Qian, H. Luo, S. Yang, N. Zhao
University of Chinese Academy of Sciences, Beijing 100049, P. R. China

J. Xu
Shenzhen University, Shenzhen 518061, P. R. China
E-mail: jxu@iccas.ac.cn

L. Pattelli, D. S. Wiersma
European Laboratory for Non-Linear Spectroscopy (LENS), Università di Firenze, Sesto Fiorentino, FI 50019, Italy
E-mail: l.pattelli@inrim.it

L. Pattelli, D. S. Wiersma
Istituto Nazionale di Ricerca Metrologica (INRiM), Turin 10135, Italy

Energy-saving cooling materials with strong operability are desirable towards sustainable thermal management. Inspired by the cooperative thermo-optical effect in fur of polar bear, we develop a flexible and reusable cooling "skin" via laminating a polydimethylsiloxane film with a highly-scattering polyethylene aerogel. Owing to its high porosity (97.9%) and tailored pore size of 3.8 ± 1.4 μm, superior solar reflectance ($\bar{R}_{sun}$ ~0.96) and high transparency to irradiated thermal energy ($\bar{\tau}_{PE,MIR}$ ~0.8) can be achieved at a thickness of 2.7 mm. Combined with low thermal conductivity (0.032 W m$^{-1}$ K$^{-1}$) of the aerogel, the cooling "skin" exerts midday sub-ambient temperature drops of 5–6 °C in a metropolitan environment, with an estimated limit of 14 °C under ideal service conditions. We envision that this generalized bilayer approach will construct a bridge from night-time to daytime radiative cooling and pave the way for economical, scalable, flexible and reusable cooling materials.

High temperature in extreme weather nowadays increasingly threatens human health and economy by inducing hyperthermia, materials aging and even causing fires. [1-3] To mitigate the negative effects of excessive heat, economical, scalable and energy-saving cooling materials are desirable. In nature, living beings have developed efficient strategies harnessing both optical and thermal management to limit the consumption of internal energy. [4-7] One prominent example is the polar bear, which uses its fur to retain the heat by a combination of radiative approach (transmit sunlight to warm the black skin while reflecting MIR radiation from the body) and non-radiative approach (thermal insulation from air convection) (**Fig. 1**a-i).[4, 8-10] This raises the interesting question whether it would be possible to design a flexible cooling "skin" exploiting this synergistic thermo-optical effect to achieve the opposite goal, i.e., to reflect solar energy and radiate heat to outer space through the atmospheric window (MIR, 8–13 μm wavelength) via the passive daytime radiative cooling (PDRC) approach. [11, 12]

Despite emerging breakthroughs in toughness, machinability and scalability, [12-15] PDRC devices still suffers from limited material-selection, modest cooling performance and low operability. Strong dependence on spectral selectivity increases the difficulties to engineer the photonic band gap of emitters[16-18], eliminating most broad-band emitters capable of night-time cooling unless shielded from direct sunlight and surrounding thermal radiation[19, 20]. Regarding general ways to increase solar reflectance ($\bar{R}_{sun}$), easy-corroded metal coats lose efficacy rapidly, and large contents of white ceramic additives eventually degrade UV reflection, MIR radiation and flexibility of cooling materials. [3, 13, 21, 22] Finally, even if the ideal optical properties were fulfilled, PDRC efficiency is inevitably restrained by heat exchange with the external air[14, 20, 23, 24]. Recently, high-porosity materials covering on emitters have been reported to effectively reduce the parasitic heat gain, yet present weak mechanical strength and low emitter-independent scattering efficiency of sunlight. [23, 24]

Herein, we propose a general approach combining a highly scattering layer of polyethylene (PE) nanoflake aerogel and a commercial polydimethylsiloxane (PDMS) film laminated together to obtain a flexible cooling "skin" that can be used on various substrates repeatedly (Fig. 1a-ii). By optimizing the size and distribution of inner nanoflakes, we obtained a PE aerogel endowed with a high $\bar{R}_{sun}$ ~0.96 and MIR transmittance ($\bar{\tau}_{PE,MIR}$) ~0.8 at a thickness of just 2.7 mm, showing a largely superior scattering efficiency as compared to other similar materials proposed for PDRC[3, 24, 25]. In addition to the extreme low mass of PE (97.9% porosity), a tailored pore diameter ($D_{pore}$) of 3.8 ± 1.4 μm also contributes to MIR transmission by favouring forward scattering through the whole atmospheric window. Further, PDRC efficiency of PE/PDMS film is improved considerably by the thermal insulation from aerogel (reaching a thermal conductivity as low as 32 mW m$^{-1}$ K$^{-1}$). As a result, we demonstrated a sub-ambient temperature reduction ($\Delta T_{cool}$) of 5–6 °C measured at solar irradiance ($I_{sun}$) beyond 1000 W m$^{-2}$ in an urban area with a calculated limit of 14 °C under ideal service conditions, which is comparative or superior to other reported results[12–15, 18–21, 26, 27] while allowing for a much broader freedom in the emitter selection. By taking advantage of the enhanced solar reflectance of the PE nanoflake aerogel and its cooperative thermo-optical effect, our approach opens a promising avenue for PDRC with enriched materials and strong operability under adverse service environments.

In contrast with previous approaches based on bulk gelation or interface mediated self-assembly, [28, 29] we have developed our PE nanoflake network (Fig. 1b) starting from a bi-continuous structure comprising PE-rich and paraffin-rich phases in thermally induced phase separation (TIPS, Fig. 1c and Supplementary note 1). High molecular weight PE ($\bar{M}_\eta$ ~4.5×10$^6$) and paraffin wax (melting temperature of 48–50 °C) are preferred to prepare aerogels for two reasons. First, the enormous kinetic asymmetry (difference in the mobility of PE and paraffin molecules) retards the phase growth of bi-continuous structure and retains the paraffin-rich phases in smaller sizes when fixed by PE crystallization (Fig. 1c-v).[30, 31] Thereby the pores formed by PE nanoflakes, which act as light-scattering elements, can be tailored and distributed with a fine-tuned density after paraffin removal. Second, using paraffin wax instead of more commonly used solvents such as liquid paraffin, leads to a gel with significantly superior shape machinability and dimensional stability (Supplementary note 2).

Thanks to these improved structural stabilities, the as-prepared PE nanoflake networks exhibit porosities as high as 99.4% (Supplementary note 3) and low thermal conductivity varying accordingly in the range between 22–32 mW m$^{-1}$ K$^{-1}$. In the following, we consider a sample with 97.9% porosity which shows excellent mechanical strength (Compression modulus is 0.72 ± 0.05 MPa, Supplementary Fig. 9) as well as high reflectance against solar irradiation ($\bar{R}_{sun}$ ~0.96) and transparency through the atmospheric window ($\bar{\tau}_{PE,MIR}$ ~0.8) (Fig. 1d). These properties are qualitatively illustrated in Figure 1d, showing a swan-shaped aerogel at a lower temperature than the background fabric under direct sunlight (Fig. 1d-i), and its transparency to MIR radiating from the human hand at room temperature (Fig. 1d-ii).

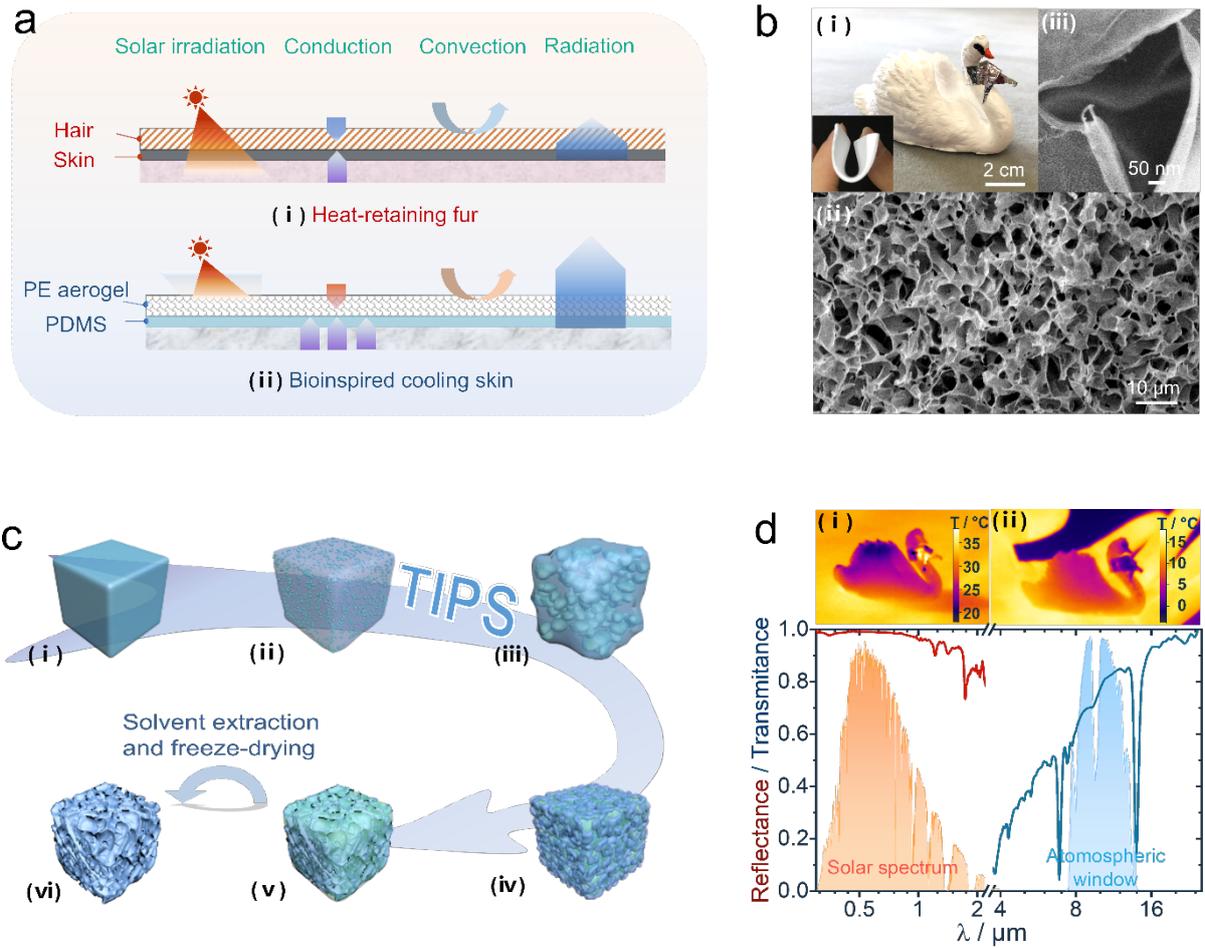

**Figure 1.** Bioinspired bilayer route with highly-scattering aerogel. a) Schematic photothermal management of heat-retaining fur of polar bear (i) and bioinspired cooling "skin" with PE aerogel of 97.9% porosity (ii, b–d). b) Photographs of the swan-shaped aerogel decorated with multi-wall carbon nanotubes coating (black), acrylic painting (orange) and aluminium foil stickers (silver) (i) and its morphology under scanning electron microscope (SEM) (ii–iii); inset of i shows flexibility of the aerogel. c) Structural evolution during the preparation of aerogel: homogeneous PE/paraffin solution (i); island morphology induced by the precipitation and growth of paraffin-rich phase (ii, iii); bi-continuous networks of PE-rich and paraffin-rich phases (iv); gel physically crosslinked by the crystallization of PE (v); PE nanoflake network after paraffin removal by solvent extraction and freeze-drying (vi). d) Spectrum of a 2.7 mm thick aerogel; insets show thermal images of the swan-shaped aerogel shot under direct sunlight outdoor (i) and held on hands indoor (ii).

Up to date, modulating polymer-based structures to achieve sufficient $\bar{R}_{sun}$ is still a challenge, especially when the filling ratio ($f$) of PE is limited to ensure high $\bar{\tau}_{PE,MIR}$.[3, 24, 25] We found that the structures of PE aerogels are rather sensitive to the polymer concentration (1–4 wt%) (**Fig. 2**a and Supplementary Note 3), changing from loose (i), dense (ii) to spherically aggregated nanoflake networks (iii, iv). Uniaxial compression (Supplementary Fig. 10 and Supplementary note 4) was performed to investigate the scattering efficiencies of different structures using the transport mean free path $l_t$ as a figure of merit (calculated at 500 nm wavelength, see Supplementary Note 5 and 6).

Shown in Fig. 2b, for the PE aerogel prepared from 2 wt% solution ($D_{pore}$ is 3.8 ± 1.4 μm, Fig. 2a-ii), $l_t$ decreases linearly at low compression factor ($C_f \leq 5.1$, defined as the uncompressed/compressed thickness ratio) compatibly with a simple independent scattering interpretation. The shortest transport mean free path is reached at $C_f$ = 11.9. At that point, the obtained PE aerogel appears as a flexible self-standing film with $l_t$ ~2.3 μm. It is worth noting that the volume density $f$ of this compressed aerogel is only 0.24, which is significantly lower than other artificial bright candidates[35-39] prepared from materials with low refractive index (generally ranging in 1.48–1.6) (Fig. 2b-ii). At the same time, it is very close to the optimum density of other optimally scattering structures[33, 34] including that of the white scales of *Cyphochilus* and *Lepidiota stigma* beetles[34], which suggests that this value might represent an optimal density independent of the structure morphology. In general, we found that all aerogels prepared from 1-4 wt% PE solutions achieve their lowest $l_t$ when compressed by factors of 11–14 (Supplementary Fig. 12), with an absolute minimum reached by the 2 wt% sample due to its denser and highly disordered structure (Fig. 2a-ii and Fig. 2b-i).

Serving as the cover layer on top of the emitter, there is a compromise between the thickness of the aerogel and its scattering efficiency at different wavelength ($\lambda$, Supplementary Note 6 and 7). Here, we have modelled the scattering properties of the aerogel by assuming a hollow sphere as the fundamental scattering unit to calculate its corresponding scattering cross section ($\langle C_{sca} \rangle$) over the visible and MIR spectrum (see Supplementary Note 8). As can be expected, $\langle C_{sca} \rangle$ is ~1000 times smaller at 10.5 μm than at 0.5 μm wavelength (Fig. 2c), indicating the possibility to combine both high $\bar{R}_{sun}$ and $\bar{\tau}_{PE,MIR}$ by modulating the pore sizes. Interestingly, a peak of the average scattering anisotropy (factor $\langle g \rangle$) is clearly visible around $D_{pore}$ of 4 μm even after averaging over the experimentally determined polydisperse distribution of pore sizes (Fig. 2c-ii), in good agreement with the pore diameter of 3.8 ± 1.4 μm obtained for the best performing aerogel prepared at 2 wt%. In this respect, controlling the pore size distribution allows to increase significantly the transport mean free path in the MIR region while leaving the scattering efficiency at visible wavelengths unaffected (Fig. 2d).

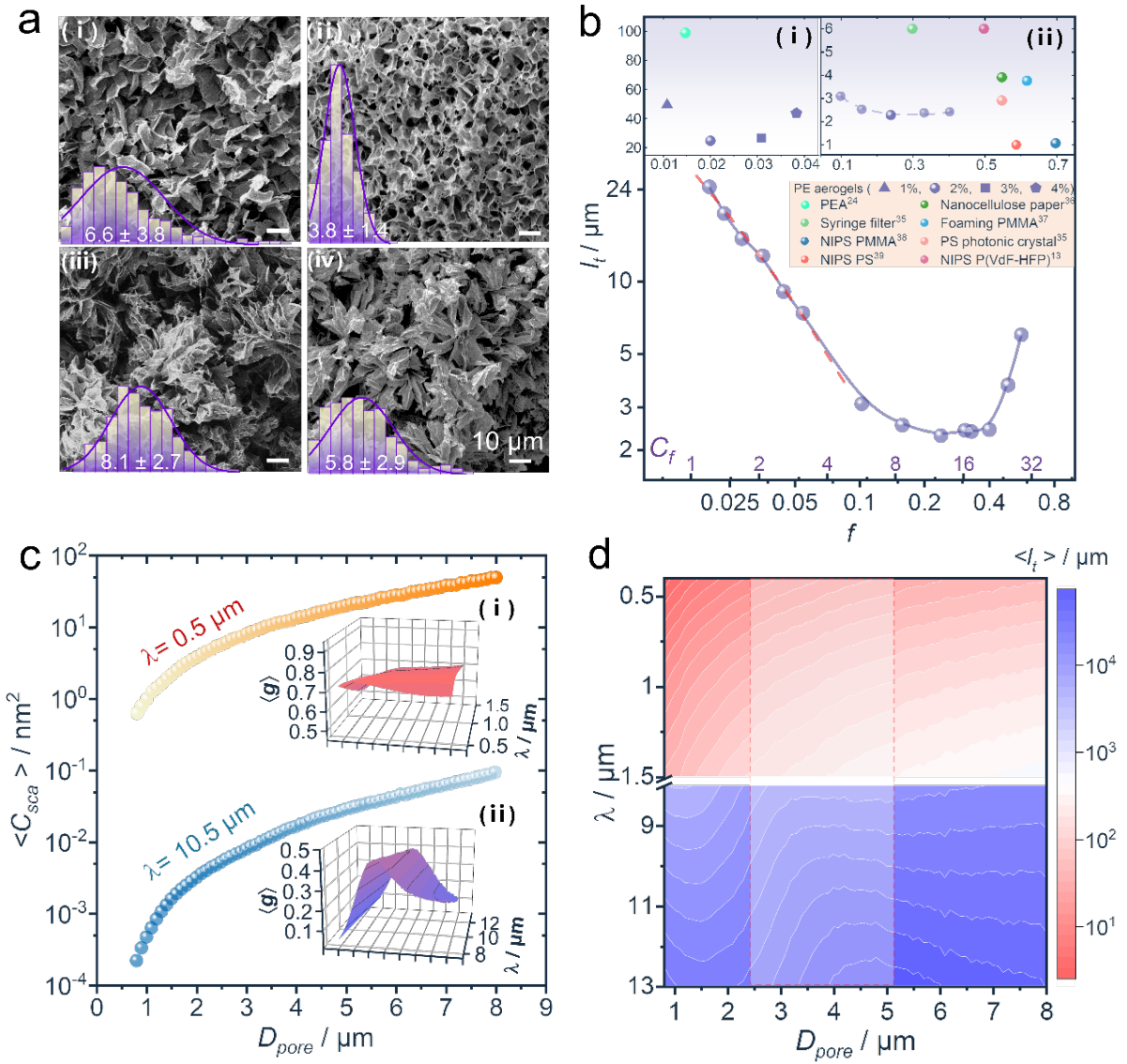

**Figure 2.** Optimizing optical structure of PE aerogel. a) SEM images of aerogels prepared from PE solutions of 1, 2, 3 and 4 wt% (i–iv), respectively, with distributions of pore sizes presented in the insets. b) Transport mean free path $l_t$ of PE aerogel under compression, with comparison among different structures shown in (a) (inset i) and other bright-white materials (inset i and ii). c) Plot of the average scattering cross section ($\langle C_{sca} \rangle$) of individual pore with different

diameters; insets are the average scattering anisotropy ($\langle g \rangle$) in the VIS-NIR (i) and MIR region (ii), respectively. d) Calculated $l_t$ of pores with different diameters.

After laminating the aerogel with a commercial PDMS layer, we obtain a flexible bilayer PDRC film with a MIR emittance ~0.8 (**Fig. 3**a). The two layers show a good adhesion strength with a peel strength at 90° estimated above 17 N m$^{-1}$ (Supplementary Note 9). Here, a PDMS film with a thickness of 150 μm was selected for its broad MIR emission, low cost, excellent flexibility and higher heat conductivity (Supplementary Note 10).[40-42] Recently, coatings have been reported to reduce the interface thermal resistance of emitters with substrate to cool[13, 43], albeit increasing the costs for their removal and reuse. On the other hand, the bilayer PDRC film is sticker-like with moderate adhesion to substrates, and can be therefore removed and/or reused to cool various materials including plastics, cements and metals by 13–27 °C (Fig. 3b). Here, the reference temperature is taken in a region covered by bare PDMS films to exclude possible errors deriving from different emittance of substrates.

The cooling efficiency of the PE/PDMS bilayer film with $\bar{R}_{sun}$ = 0.960 and $\bar{\tau}_{PE,MIR}$ = 0.788 was measured in direct contact with external air (Fig. 3c). Confronted with the adverse heat radiation from the surrounding high-rise buildings[19] (Fig. 3c, urban area in Beijing, 39.99° N, 116.32° E, altitude 58 m; continental monsoon climate), the bilayer film achieved a remarkable $\Delta T_{cool}$ of 5–6 °C under intense $I_{sun}$ exceeding 1000 W m$^{-2}$ (Fig. 3d), resulting in a measured cooling power ($P_{cool}$) of 70 ± 14 W m$^{-2}$ (Fig. 3e). In an open field with weaker $I_{sun}$ (800–865 W m$^{-2}$) and lower relative humidity (RH, below 26%) (Turpan 42.95° N, 89.18° E, altitude 23 m; continental desert climate), we observed an increased $\Delta T_{cool}$ of 6–7 °C (Fig. 3f) and a $P_{cool}$ of 78 ± 12 W m$^{-2}$ (Fig. 3g), which is still notable considered the low altitude, adverse air quality (PM10 of 74 and AQI of 63 in September 14$^{th}$; PM10 of 82 and AQI of 66 in September 16$^{th}$) and moderate insulation of the equipment (retrieved heat transfer coefficient of 3 W m$^{-2}$ K$^{-1}$ fitted from the setup in Fig. 3c, see Supplementary Fig. 19). Such service environments are generally far from the more idealized conditions where experiments have been reported in the literature[12, 13, 24].

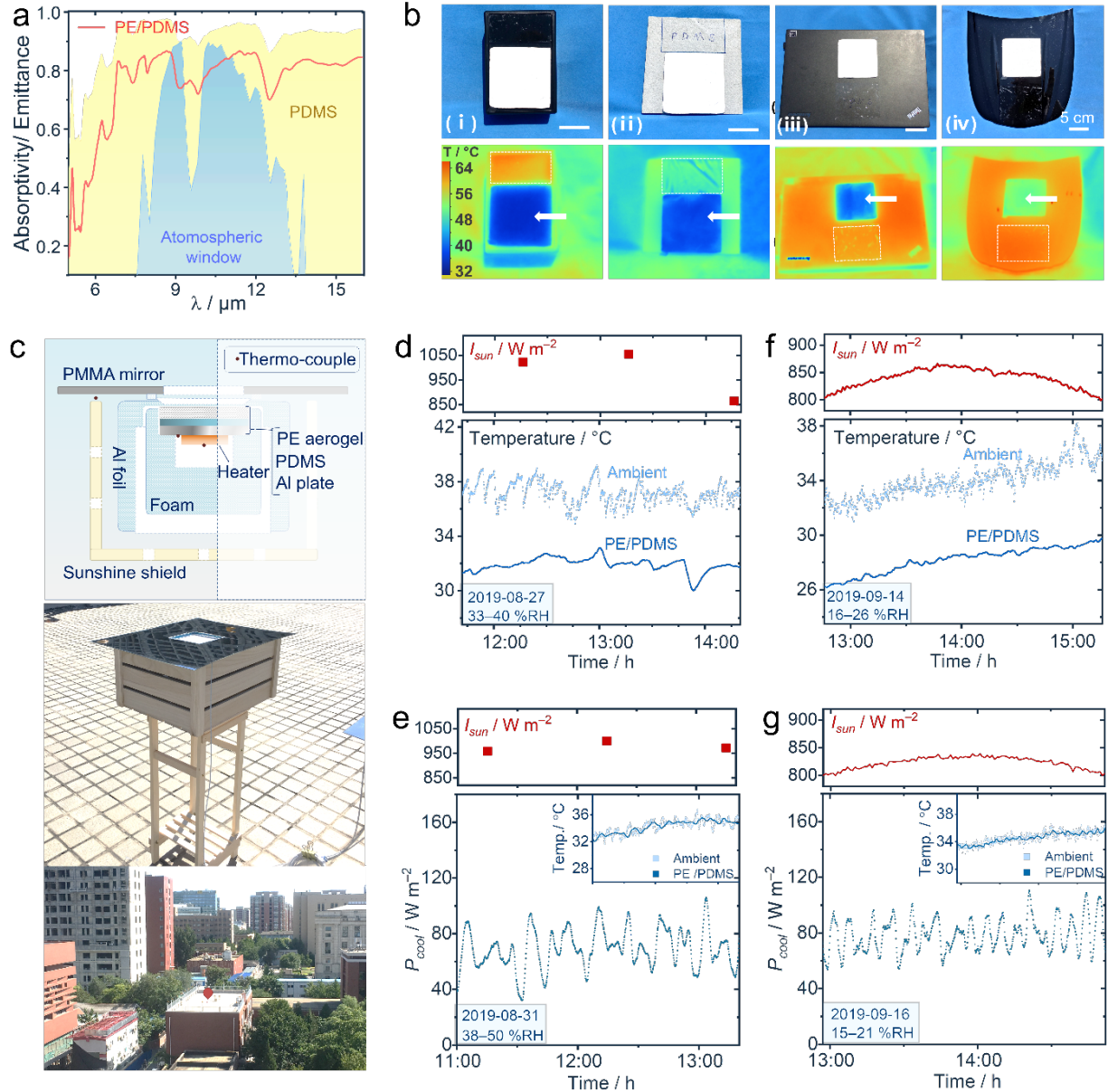

**Figure 3.** PDRC performance of PE/PDMS cooling "skin". a) MIR emission of PDMS before and after lamination with a 2.7 thick PE aerogel. b) Thermal images of the bilayer PDRC film (pointed by the white arrow) on various surfaces, including plastic packaging, 64 °C (i), cement, 52 °C (ii), laptop computer, 60 °C (iii) and front cover of a car mode, 62 °C (iv); the dashed region was covered by bare PDMS film for reference purposes. c) Schematic of the test equipment and environment in downtown Beijing. Sub-ambient temperature drops ($\Delta T_{cool}$, d and f) and cooling powers ($P_{cool}$, e and g) respectively in downtown Beijing (d and e) and an open field in Turpan (f and g) with solar irradiance ($I_{sun}$) recorded; insets show temperatures of the ambient and the PE/PDMS layer.

In order to investigate the cooperative effect between the radiative and non-radiative mechanisms on the PDRC performance, we compared PDMS layer laminated with PE aerogels at different degrees of compression $C_f$. During a gradual compression of the aerogel up to a factor of 8.6, $\bar{R}_{sun}$ and $\bar{\tau}_{PE,MIR}$ are only slightly affected (**Fig 4a**-i), while thermal resistance (estimated by the ratio of thickness to thermal conductivity, $t/k$, Fig 4a-ii) is strongly reduced determining the overall trends of sub-ambient temperature drops in simulations ($\Delta T_{cool,cal}$, Fig 4a-iii, and Supplementary Note 11). Although both $\Delta T_{cool,cal}$ and calculated cooling powers ($P_{cool,cal}$, Fig. 4b) decline with increasing compression, thermal insulation always plays an active role by virtue of the considerably lower heat conductivity of PE aerogel compared to PDMS and other PDRC materials. Specifically, the enhanced thermal insulation of the PE aerogel cover can overcome the adverse effects of weakened MIR emission of PDMS for values of $C_f < 3$, as determined by the intersection of $\Delta T_{cool,cal}$ at night when the influence of solar irradiation is excluded (Fig. 4b-inset). At this threshold ($C_f = 3$), $\Delta T_{cool}$ is ~1.5 °C lower than that obtained using uncompressed aerogel (Fig. 4c).

Compared to the bare PDMS case, a 97.9% porosity PE aerogel efficiently inhibits the nonradiative heat transfer between PDMS and hot air, thereby increasing the $\Delta T_{cool}$ by 1–2 °C at night (Fig. 4d) and 9–13 °C at noontime (Fig. 4e). Our theoretical modelling, which successfully reproduces the experimental data of Fig. 3, Fig. 4 and Supplementary Table 3, predicts a limiting cooling performance $\Delta T_{cool}$ ~14 °C for a large-area PE/PDMS film (Supplementary Fig. 20, Supplementary Note 11 and 12). It is worth noting that the PDMS emitter used alone cannot generate a net PDRC effect (Fig. 4e), which highlights the true synergistic nature of our proposed approach. Using a thicker aerogel of the achieved highest porosity (99.4%), thermal insulation will be further enhanced and may result in an impressive cooling performance,[24] while it is still hard to get rid of the dependence of an additional reflector to compensate $\bar{R}_{sun}$ or else directly at the expense of a decline of $\bar{\tau}_{PE,MIR}$ resulting in an unsatisfactory $\Delta T_{cool}$. Therefore, to extend traditional night-time cooling emitters (PDMS for instance) to PDRC, it is of vital significance to fully optimize the structure of front porous materials with a compromise among $\bar{R}_{sun}$, MIR transmittance and thermal resistance. Moreover, benefiting from the superhydrophobic characteristic of PE aerogels, dusts can be easily rinsed off to maintain the long-term performance of cooling materials (Supplementary Note. 13).

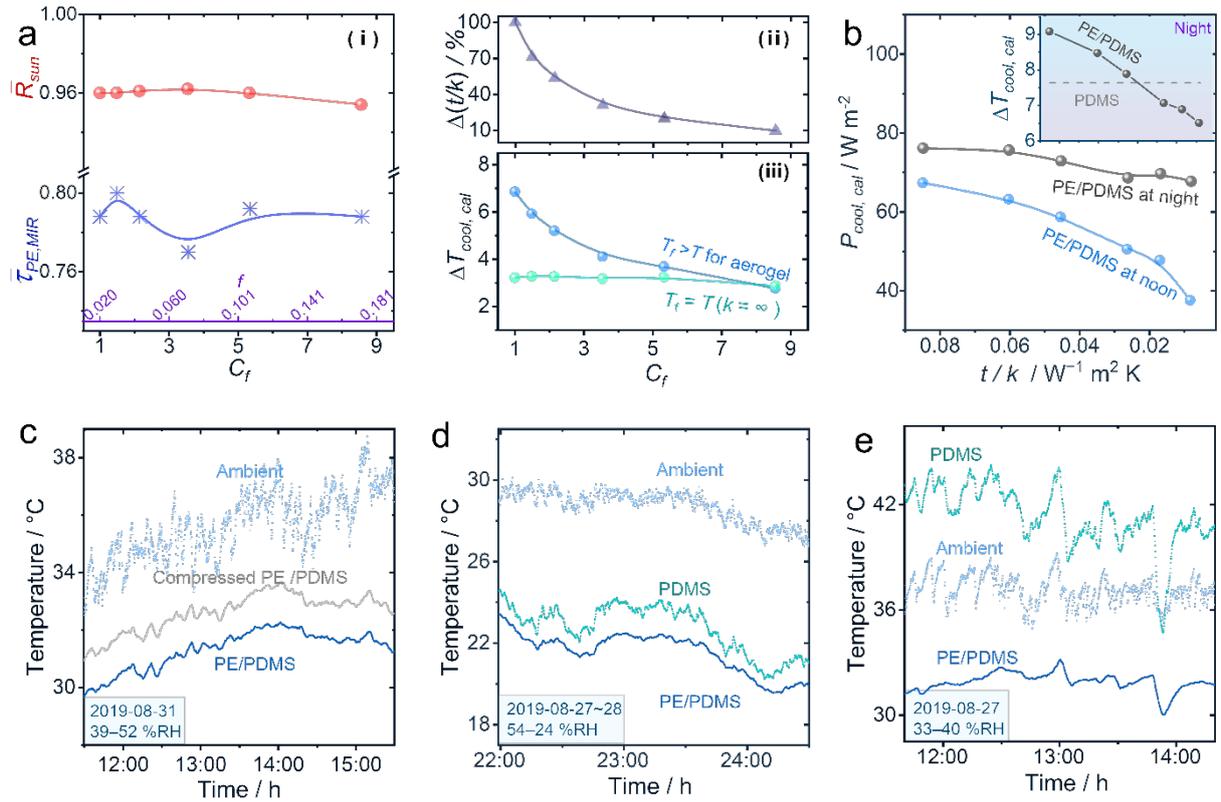

**Figure 4.** Cooperative effect between the radiative and non-radiative approaches. a) Simulated compression-dependent sub-ambient temperature drops ($\Delta T_{cool,cal}$, a-iii) using aerogel with $\bar{R}_{sun}$, $\bar{\tau}_{PE,MIR}$ (i) and thermal resistance ($t/k$, ii) change by compression, where $T_f$ and $T$ correspond to the temperature of the front and reverse side of aerogel, respectively. b) Simulated cooling powers ($P_{cool,cal}$) using aerogels with different $t/k$; inset shows their $\Delta T_{cool,cal}$ compared to bare PDMS at night. c) $\Delta T_{cool}$ of bilayer PDRC films with PE aerogel before and after compression by a factor of three. Cooling performance of PDMS covered with or without PE aerogel at (d) midnight and (e) noontime.

In summary, we demonstrated a bioinspired bilayer route lowering the barriers to PDRC with improved efficiency. A cooling "skin" integrating a flexible PE aerogel onto a simple PDMS film has been developed that can be easily applied on various substrates and used repeatedly. A PE aerogel with 97.9% porosity was shown to effectively reflect sunlight and resist to heating by the external hot air. As a result, we demonstrated a $\Delta T_{cool}$ of 5–6 °C at noontime in a metropolitan environment, with a predicted cooling limit of 14 °C under ideal service conditions. Based on the optimized scattering efficiency of PE aerogel ($\bar{R}_{sun}$ ~0.96, $\bar{\tau}_{PE,MIR}$ ~0.8 for a 2.7 mm thick layer), the synergistic approach with non-radiative effect enables radiative cooling to reach its full potential, which is a key to expand the available materials for PDRC and their applications in real life. Since bright-white aerogels can be prepared from a wide array of materials (Supplementary Note 14), we expect that this method will inspire the exploration of different coatings with optimized scattering efficiency. In addition to passive cooling, bilayer devices for photo–thermal management are possibly available by changing the energy-conversion material, allowing for abundant applications in infrared stealth, solar-steam generation and thermoelectric generation.[8, 44, 45]

**Experimental Section**

## Methods

*Preparation of PE aerogel*: Typically, 1.5 g PE powder ($\bar{M}_\eta$ ~4.5×10$^6$), 73.5 g paraffin (melting temperature of 48–50 °C) and 7.5 mg antioxidant 1010 were stirred at 150 °C for 3 hours. The resultant homogeneous solution was casted into a pre-heated mould and cooled by ice water (Supplementary Fig. 1). The as-formed gel was extracted with cyclohexane for several times, and then freeze-dried. PE aerogels of different porosities were prepared in the same way but with varied PE concentrations (0.005–4 wt%). Diverse shapes of aerogel were prepared by casting in different moulds, or by drafting or dip coating.

*Preparation of compressed aerogels*: The compression of aerogels was operated between two metal plates, as illustrated in Supplementary Fig. 2. Weighing paper was put between metal plates and samples before the compression.

*Adhesion of aerogel and polydimethylsiloxane*: An aerogel slab was put on the surface of polydimethylsiloxane film swelled by cyclohexane, and then adhered together by the negative pressure cavity due to the diffusion and volatilization of cyclohexane.

## Acknowledgements

The authors would like to acknowledge the financial support from Chinese Academy of Sciences (No. QYZDB-SSW-SLH025) and National Natural Science Foundation of China (51522308).